\documentstyle[preprint,eqsecnum,aps]{revtex}

\def\o{\overline}

\begin{document}
\draft
\preprint{CGPG-94/6-3, UCSBTH-94-39, gr-qc/9409049}
\title{Observables and a Hilbert Space for Bianchi IX}
\author{Donald Marolf\cite{Marolf}}
\address{Physics Department, The University of California,
Santa Barbara, California 93106} \date{September 1994}
\maketitle

\begin{abstract}
We consider a quantization of the Bianchi IX cosmological model
based on taking the constraint to be a self-adjoint operator in an
auxiliary Hilbert space.  Using a WKB-style self-consistent
approximation, the constraint chosen is shown to have
only continuous spectrum at zero.  Nevertheless, the auxiliary
space induces an inner product on the zero-eigenvalue
generalized eigenstates such that the resulting physical Hilbert
space has countably infinite dimension.
In addition, a complete set of
gauge-invariant operators on the physical space is constructed
by integrating differential forms over the spacetime.
The behavior of these operators indicates that this
quantization preserves Wald's classical result that the
Bianchi IX spacetimes expand to a maximum volume
and then recollapse.

\end{abstract}
\pacs{ICP:0460}

\section{Introduction}
\label{intro}

Finite dimensional cosmological models are a mainstay of modern
general relativity for the simple reason that they
address many exciting conceptual issues without forcing the
researcher to confront the complicated, infinite dimensional, and in
the quantum context perturbatively non-renormalizable formalism of
full 3+1 gravity.
Our interest here will be the quantum theory and our
motivation for studying this model is exactly that stated above.
Again, we ask the historical questions (e.g., \cite{tril,KK,Kuchar,QORD}):
``What are the quantum observables?" ``How are they constructed or
even defined?"  And, ``On what Hilbert space, if any, do they act?"

A proposal for addressing these questions was introduced in
\cite{QORD} and applied to finite dimensional parameterized Newtonian
systems, the free relativistic particle,
and ``separable, semi-bound" models (which include the
Kantowski-Sachs and locally rotationally
symmetric (LRS) Bianchi IX cosmological models, see \cite{QORD}).  For such
systems, the approach was found to define a Hilbert space of physical
states, to produce a complete set of gauge-invariant
quantum operators, and to reproduce (in a certain sense) the
classical feature that the semi-bound models describe spacetimes with
homogeneous spatial hypersurfaces first expand and then
recollapse.  The strategy was based on that of DeWitt \cite{Bryce}
and takes seriously the idea that observables in General Relativity
are given by integrals of differential forms over the
spacetime manifold.  The analogues of such integrals were, in
fact,  performed in the quantum theory.

In order to implement this basic proposal, a mathematical
framework must first be introduced.  The main framework of
\cite{QORD} is based on that of Dirac \cite{PAMD} and
considers an auxiliary Hilbert space in which
gauge-dependent operators act.  For the models in question,
a physical inner product was induced on the generalized
eigenvectors of the constraint by this auxiliary structure
{\it despite} the fact that no solution of the constraint was
normalizable in the auxiliary inner product.

However, \cite{QORD} considered only relatively simple models.
While some models addressed by section V of \cite{QORD} are
not explicitly solvable even classically,
the usual minisuperspace models (Bianchi I,
Kantowski-Sachs, and LRS Bianchi IX) that fall into this
category are solvable and have even been deparameterized \cite{atu}.
Also, the separable nature of these systems was explicitly used in
\cite{QORD} to construct the physical Hilbert space and to
study convergence of the integrals that define the gauge invariant operators.

Our goal here is to apply these same ideas to the more complicated
Bianchi IX, or Mixmaster, minisuperspace \cite{Misner,BKL,MTW}.
Like other Bianchi
models, it describes a class of solutions to the vacuum Einstein
equations that are foliated by a family of homogeneous spacelike
hypersurfaces.  This preferred family effectively reduces the
gauge symmetry of full General Relativity to that of time
reparameterizations.

Unlike the simpler models, the Bianchi IX cosmology has not
been completely solved or deparametrized classically.  In fact, it
is often \cite{chaos,DEQ} considered to be an example of chaotic
behavior in General Relativity.  It therefore constitutes a much
more stringent test of the proposal of \cite{QORD}.
We shall see that, by introducing new tools and a WKB-style
self-consistent approximation, all of the corresponding
results are obtained.

We begin our presentation by reviewing the method of
\cite{QORD} in section \ref{prelim}.
Section \ref{IX}
then applies these techniques to the mixmaster model, beginning with a
description of the quantum theory (section \ref{descr}) and
followed by a study of solutions to our constraint (\ref{sols}) and
a characterization of the physical Hilbert space (\ref{Hphys}).
One consequence of our results is that,
within the validity to this approximation, we show that the space of solutions
to the Hamiltonian constraint for Bianchi IX is infinite dimensional.
Finally, a complete set of
gauge-invariant operators is constructed and the
recollapsing behavior of our quantum theory is verified in \ref{ops}.
We close with a summary
of the results and a comparison with other work.

\section{The General Approach}
\label{prelim}

Because our goal is to apply the techniques of \cite{QORD} to the
Bianchi IX
minisuperspace, we now summarize the relevant points of
\cite{QORD} for the reader's convenience.
The general setting considered is that of a time
reparameterization invariant system with Hamiltonian constraint:
\begin{equation}
h = 0
\end{equation}
for some function $h$ on the phase space $\Gamma = {\bf R}^{2N}$
(see \cite{QORD} for a more general treatment).
Here, as in \cite{QORD}, we will take capital letters to stand
for quantum operators and lower case latin letters to refer to their
classical counterparts.  For each coordinate $q_i,p_j$ on $\Gamma$,
a family $Q_i(t),P_j(t)$ of self-adjoint
operators is introduced on an (auxiliary) Hilbert space ${\cal H}
= L^2({\bf R}^N)$.

Now, choose some time independent symmetric factor ordering
$H(Q_i(t),P_j(t))$ corresponding to $h(q_i,p_j)$ and consider the
associated self-adjoint operator $H$.  The families
$Q_i(t)$, $P_j(t)$ are chosen to satisfy
\begin{equation}
\label{fam}
{{dA} \over {dt}} = i [N(t) H,A]
\end{equation}
for any $A = A(Q_i(t),P_j(t))$ and a family of lapse operators $N(t)$
that are proportional to the
identity on ${\cal H}$; $N(t) = n(t)\openone$.
The commutators of the families form a quantized version of a
generalized Peierls bracket \cite{gp} and the
formulation of \cite{QORD} provides a notion of gauge
transformations corresponding to time reparameterizations.

This structure is to be applied toward two goals:  $i$) The
construction of a physical Hilbert space ${\cal H}_{phys}$
and $ii$) The construction of a complete set of gauge-invariant
operators that commute with $H$.  Here, we use the same notion of
a complete set as in \cite{QORD}.  That is, we say that a set of
operators on ${\cal H}_{phys}$ is complete if the corresponding
set of phase space functions obtained in the classical limit
separates the points of the phase space.
For parameterized
Newtonian systems and so-called ``separable semi-bound" systems (which
include the Kantowski-Sachs and LRS Bianchi IX cosmological models),
both of these objectives are attained in \cite{QORD}.  We will now
see that they may be accomplished for the full-fledged Bianchi IX
model as well.

The physical Hilbert space is to be defined as follows.  We first
find a convenient ``energy basis" of states $|n,E\rangle$ for
$E \in [\lambda,0]$ (for some $\lambda < 0$) and $n \in {\bf Z}^+$
that satisfy $H|n,E\rangle = E|n,E\rangle$ and
\begin{equation}
\langle n',E'|n,E\rangle = \delta(E-E') \delta_{n,n'}
\end{equation}
and span the subspace of ${\cal H}$ corresponding to the interval
$[\lambda,0]$ in the continuous spectrum of $H$.  The physical
space ${\cal H}_{phys}$ is to be the closure of the vector space
generated by
the symbols $\{|n,0\rangle \}$ for $n \in {\bf Z}^+$ with
the physical inner product: $(|n',0\rangle, |n,0\rangle)_{phys}
= \delta_{n,n'}$.

For a gauge-invariant
operator $A$ on ${\cal H}$ that commutes with $H$, we
would then like to consider an operator $A_{phys}$ on ${\cal H}_{phys}$
given by:
\begin{equation}
A_{phys}|n,0\rangle  = \sum_k a_{nk}(0) |k,0\rangle
\end{equation}
where the coefficients
$a_{nk}$ are defined by $A|n,E\rangle = \sum_{k}a_{nk}(E)
|k,E\rangle$.  This procedure is satisfactory when $a_{nk}(E)$
is continuous in $E$ on $[\lambda,0]$ and in this case
$A_{phys}$ is a symmetric bilinear form on ${\cal H}_{phys}$ when $A$
is symmetric on ${\cal H}$.  In this way our physical
inner product captures the classical reality conditions \cite{AA} to the
same extent as the auxiliary inner product on ${\cal H}$.

Such gauge invariant operators were constructed in \cite{QORD} by
integrating over the parameter $t$
that labels the families of operators
that appear in \ref{fam}.
In particular,
given two operators $A$ and $Z$ on ${\cal H}$, we follow
\cite{QORD} and consider
\begin{equation}
\label{lobs}
[A]_{Z=\tau}^L = \int_{-\infty}^{\infty} dt A(t) ({{\partial}
\over {\partial t}} \theta(Z(t) - \tau))
\end{equation}
and
\begin{equation}
\label{robs}
[A]_{Z=\tau}^R = \int_{-\infty}^{\infty} dt ({{\partial}
\over {\partial t}} \theta(Z(t) - \tau)) A(t)
\end{equation}
where $\theta(Z(t) - \tau)$ is the projection onto the non-negative
spectrum of $Z(t) - \tau$.  The object
\begin{equation}
\label{obs}
[A]_{Z= \tau} = {1 \over 2} ( [A]^L_{Z = \tau} + [A]^R_{Z = \tau})
\end{equation}
is formally symmetric and in the classical limit is simply a
function on the space ${\cal S}$ of solutions that, when applied
to a particular solution $s$, is a linear combination of the values of
$A$ at the points along $s$ at which $Z = \tau$.  Furthermore,
when the integrals \ref{lobs} and \ref{robs}  converge, they
define operators that commute with $H$\cite{QORD}.  For appropriate
systems and choices of $A$ and $Z$, this procedure was
shown in \cite{QORD} to produce a complete set of well-defined operators
$([A]_{Z = \tau})_{phys}$ on ${\cal H}_{phys}$.  Because the
correspondence between the gauge invariant
$A$ and the physical operator $A_{phys}$ is direct, we
now drop the subscript $phys$ from these operators.

One final property derived in \cite{QORD} is that a quantization of
this type captures the classical ``recollapsing behavior" of
separable semi-bound
cosmological models.  That is,
in contrast to quantization schemes based on the Klein-Gordon
inner product \cite{KG}, there is a quantum analogue of
the classical result that
these models expand and then recollapse.
For such models, if $Z$ is taken to be
the scale factor $\alpha = \ln(\det g)$ for the metric $g$ on a
homogeneous spatial hypersurface, then
the operators $[A]_{\alpha = \tau}$
vanish in the large $\tau$ limit.
Because the corresponding classical quantities
satisfy $[a]_{\alpha = \tau} \rightarrow 0$ as
a direct consequence of the fact that a given
solution $s$ will not reach arbitrarily large values of $\alpha$, we
interpret
\begin{equation}
\label{rec}
[A]_{\alpha = \tau} \rightarrow 0 \ \rm{as} \ \tau \rightarrow +\infty
\end{equation}
as the quantum statement of recollapse.
The Bianchi IX model also recollapses classically \cite{Wald} and we
shall see that, when quantized as
above, it recollapses quantum mechanically
as well.

\section{The Quantization of Bianchi IX}
\label{IX}

The discussion of \cite{QORD} considered the parameterized Newtonian
particle and separable semi-bound
models, but our interest here is in
the more complicated Bianchi IX minisuperspace.  We will see that the
same (qualitative) results can be obtained as we will now
construct $i$) a non-trivial physical Hilbert space from generalized
eigenvectors of the constraint operator
as well as $ii$) a complete set of
gauge invariant operators and $iii$)
verify that our quantum theory leads
to the recollapsing behavior derived
classically in \cite{Wald}.

We follow much the same path as in section V of \cite{QORD}.
Section \ref{descr} presents the system and sections \ref{sols} and
\ref{Hphys}
construct a convenient ``energy" basis
for the auxiliary space ${\cal H}$ in terms of which the
physical states can be simply characterized.  This part of
the analysis is carried out at the somewhat physical
level of WKB-style self-consistent approximations.
Section \ref{ops} considers
operators \ref{obs} of the form $[A]_{\alpha = \tau}$ for
$\alpha = \ln (\det g)$ and $g$ the three-metric on such a slice and
finds that matrix elements of these operators
converge for a complete set of
appropriate $A$ so such
$[A]_{\alpha = \tau}$ exist as bilinear forms on
${\cal H}_{phys}$.  We then modify
the construction of $[A]_{\alpha = \tau}$
by introducing a ``level operator"
$L$ whose physical interpretation is
less clear, but which improves the behavior of our operators.
The modified $\widetilde{[A]}_{\alpha = \tau}$ can then be
proven to define (for appropriate $A$) a complete set of {\it bounded}
operators on ${\cal H}_{phys}$.
Both the forms $[A]_{\alpha = \tau}$ and the operators $\widetilde{
[A]}_{\alpha = \tau}$ vanish as $\tau \rightarrow \infty$,
demonstrating that our model
preserves the classical recollapsing behavior.

\subsection{The Quantum System}
\label{descr}

The mixmaster model describes anisotropic
homogeneous solutions to the
$3+1$ vacuum Einstein equations.  For this case, the metric $g_{ij}$
on a homogeneous slice \cite{MTW} may be written in the form
\begin{equation}
\label{ma}
g_{ij} = e^{2 \alpha} (e^{2\beta})_{ij}
\end{equation}
for a matrix $\beta_{ij}$ given by $diag(\beta^+ + \sqrt{3} \beta^-,
\beta^+ - \sqrt{3} \beta^-, -2\beta^+)$ which defines the
anisotropy parameters $\beta^{\pm}$ and the
scale factor $\alpha$.  Taking the metric to be in this
form fixes all of the gauge
freedom except that of time reparameterizations.

Since the classical phase space is described
by $(\alpha, \beta^+, \beta^-, p_{\alpha}, p_{\beta^+}, p_{\beta^-})
\in {\bf R}^6$ we take the auxiliary space
${\cal H}$ to be $L^2({\bf R}^3,d^3x)$ following \cite{QORD}.
Our quantum theory will be built from the configuration
operators $\alpha, \beta^{\pm}$ which act on functions $\psi \in
L^2({\bf R}^3)$ by
multiplication as well as the derivative operators
$P_{\alpha} = -i\hbar {{\partial}
\over {\partial \alpha}}$, $P_{\beta^{\pm}} = -i \hbar {{\partial}
\over {\partial \beta^{\pm}}}$ which are the corresponding momenta.

Recall that the full Hamiltonian
constraint for this model may be
written \cite{MTW} as
\begin{equation}
\label{IX constr}
H = - P^2_{\alpha} + P^2_{\beta^+} +
P^2_{\beta^-} + e^{4 \alpha}(V(\beta) -1 )
\end{equation}
using a rescaled lapse $N(t)
= ({{3 \pi} \over 2})^{1/2} e^{-3 \alpha} \tilde{N}(t)$
and Hamiltonian $H =
({2 \over {3 \pi}})^{1/2} e^{3 \alpha} \tilde{H}$.
(where $\tilde{N}$ and
$\tilde{H}$ are the usual ADM lapse and Hamiltonian
of general relativity \cite{ADM} applied to the minisuperspace
ansatz \ref{ma}).  Note that by
writing \ref{IX constr} as a quantum operator, we
have chosen a certain factor ordering of the classical expression.  The
explicit form of the potential is given by
\begin{equation}
\label{V}
V(\beta) = {1 \over 3}e^{-8 \beta^+} -{4 \over 3} e^{-2 \beta^+}
\cosh 2 \sqrt{3} \beta^- + 1 +{2 \over 3} e^{4 \beta^+}
(\cosh 4 \sqrt{3} \beta^- -1)
\end{equation}
which has a global minimum $V=0$ at the origin $\beta^{\pm} = 0$
but has no other critical points.
The operator \ref{IX constr} is symmetric on $L^2({\bf R}^3)$ and
we shall assume that some suitable extension is in fact self-adjoint.
This extension will also be written as $H$ and no distinction will be
made between the two operators.

The operator $H$ of \ref{IX constr} is Hermitian and we use it
to introduce a ``time dependent" set of
configuration bases $\{|\alpha, \beta^+,\beta^-;t'\rangle \}$ which are
related through
\begin{equation}
|\alpha, \beta^+,\beta^-;t'\rangle = e^{i(t'-t)H}|\alpha,\beta^+,
\beta^-; t \rangle.
\end{equation}
The (time dependent)
direct product
decomposition ${\cal H} = {\cal H}_{\alpha;t}
\otimes {\cal H}_{\beta;t}$
induced by the factorization $|\alpha, \beta^+,\beta^-;t\rangle
= |\alpha;t\rangle \otimes |\beta^+,\beta^-;t\rangle = |\alpha;t
\rangle |\beta^+,\beta^-;t\rangle$
will be used heavily in what follows.

As our system is time reparametrization invariant and \ref{IX constr} is
classically constrained to vanish, it should be emphasized that this $t$
merely serves as a label and is not the Newtonian time parameter of
non-relativistic quantum mechanics.  Nevertheless, it behaves similarly
when used as a technical tool and we use it to construct
the families $Q_i(t)$ and $P_j(t)$.  These will be
defined by $Q_i(0) = Q_i$, $P_j(0)=P_j$
and
\begin{eqnarray}
Q_i(t) &=& \exp\Big( i H \int_{0}^t dt' N(t') \Big) Q_i \Big(
-i H \int_0^t dt' N(t') \Bigr) \cr
P_j(t) &=& \exp\Big( i H \int_{0}^t dt' N(t') \Big) P_i \Big(
-i H \int_0^t dt' N(t') \Bigr)
\end{eqnarray}
for some $N(t) = n(t) \openone$, $n(t)\in {\bf R}$.  This ``lapse
operator" and its relation to gauge transformations are discussed in
\cite{QORD}.

The generalized eigenvectors of the constraint \ref{IX constr}
with eigenvalue zero will form the physical Hilbert space
${\cal H}_{phys}$.
In order to study solutions of $H|\psi\rangle = 0$,
 we will consider the operator
\begin{equation}
H_1 = P^2_{\beta^+} + P^2_{\beta^-} + e^{4 \alpha}(V(\beta) -1 )
\end{equation}
and for each $t'$, introduce a family $H^*(\alpha;t')$ of operators
on ${\cal H}_{\beta;t'}$ parametrized by $\alpha$:
\begin{equation}
H_1 = \int d \alpha |\alpha;t' \rangle \langle \alpha;t' | \otimes
H^*(\alpha ; t').
\end{equation}
This $H^*(\alpha;t')$ is a
Hermitian operator on the Hilbert space
${\cal H}_{\beta;t'} = L^2({\bf R}^2)$ of square integrable
functions of $\beta^+$ and $\beta^-$.  We similarly define
$P^*_{\beta^{\pm}}$ and $\beta^{\pm *}$.
Note that, for various $\alpha$, the $H^*(\alpha)$ differ only in the
overall scale of the potential $V^*(\beta^{\pm*};\alpha) = e^{4 \alpha}
(V(\beta^{\pm*}) -1)$ and that, because the region in which
the potential is less than $E$ for any $E \in {\bf R}$ has finite
area (see appendix A),
the spectrum of these operators is entirely discrete \cite{QM}
and the degeneracy of each spectral projection is finite.

Thus, we may label the eigenstates of $H^*(\alpha;t')$ in
${\cal H}_{\beta;t'}$ as
$|n ; \alpha;t' \rangle_{\beta}$ for $n\in {\bf Z}^+$.
The state
$|n ; \alpha;t'\rangle_{\beta}$ is to have the eigenvalue $E_n(\alpha)$
(which does not depend on $t'$) of $H^*(\alpha;t')$ and these
states are orthonormal:
${}_{\beta}\langle n';\alpha;t'|n;\alpha;t'\rangle_{\beta}
 = \delta_{n,n'}$.  Since the family $H^*(\alpha;t')$ is analytic in
$\alpha$,
the states $|n ; \alpha; t\rangle $ may also be chosen smooth in the label
$\alpha$ for each $n$ \cite{RSIV} and
we may express any state $|\psi\rangle$
in ${\cal H}$ as
\begin{equation}
\label{psidef}
|\psi\rangle = \int d \alpha \sum_n f_n(\alpha) |\alpha\rangle
|n ; \alpha \rangle_{\beta}
\end{equation}
for some $f_n(\alpha)$.
Until \ref{ops}, we will work at some fixed
value of $t'$ and suppress this label on
the states, operators, and spaces.

\subsection{Solutions of the Constraint}
\label{sols}

In the following, we operate at the somewhat physical level of
self-consistent approximations and
to use \ref{psidef} to study the
eigenvectors of $H$; that is, to solve $H|\psi \rangle = E |\psi\rangle$.
We begin by considering
the action of $H$ on \ref{psidef}:
\begin{equation}
\label{expand}
H | \psi \rangle = \int d \alpha \sum_n  \Bigl(\hbar^2
{{\partial}^2 \over {\partial \alpha^2}} \bigl( f_n(\alpha) |n;
\alpha \rangle_{\beta} \bigr)
+ f_n(\alpha) E_n(\alpha) |n ; \alpha\rangle_{\beta} \Bigr) |\alpha\rangle
\end{equation}
Now, from the equation
\begin{equation}
\label{baeig}
(H^*(\alpha) - E_n(\alpha)) | n ; \alpha \rangle_{\beta} = 0
\end{equation}
in ${\cal H}_{\beta}$
it follows that the states $|n ;  \alpha \rangle_{\beta}$
may be chosen to satisfy
\begin{eqnarray}
{{\partial} \over {\partial \alpha}} |n ; \alpha \rangle_{\beta}
&=& {1 \over {(H^*(\alpha) - E_n(\alpha))_{\perp}}}
(-4 e^{4 \alpha} (V-1)) |n ; \alpha \rangle_{\beta}
\cr &=& \sum_m  M_{mn} |m ; \alpha \rangle_{\beta}
\end{eqnarray}
where $ {1 \over {(H^*(\alpha) - E_n(\alpha))_{\perp}}} $ is the
inverse of $H^*(\alpha) - E_n(\alpha)$ except when acting on
states $|m;\alpha\rangle_{\beta}$
with $E_m(\alpha) = E_n(\alpha)$ (i.e., states
degenerate with $|n;\alpha\rangle_{\beta}$),
which it annihilates.

The matrix $M_{mn}$ is antihermitian since
$M_{nm}$ vanishes when $E_n = E_m$ and, for $E_n \neq E_m$, we have
\begin{eqnarray}
M_{mn} &=& \langle m ; \alpha |
{{-4}
\over {(H^*(\alpha) - E_n(\alpha))_{\perp}}}  e^{4 \alpha} (V-1)
|n ; \alpha \rangle
\cr &=&
{ {\langle m; \alpha | -4 e^{4 \alpha} (V-1) | n; \alpha \rangle} \over {
E_m(\alpha) - E_n(\alpha) }}
\end{eqnarray}
for which the numerator is Hermitian and the denominator is real
and antisymmetric.
We now use $M_{nm}$ to rewrite the derivative in \ref{expand} as
\begin{equation}
{{{\partial} \over {\partial \alpha}}}
( \sum_n f_n(\alpha) |n; \alpha  \rangle_{\beta})
= \sum_n ( {{{\partial} \over {\partial \alpha}}}
f_n(\alpha) + \sum_m f_m(\alpha) M_{nm}) |n ; \alpha \rangle_{\beta}.
\end{equation}

Our goal is to study generalized eigenvectors of $H$ for which
the solutions
$|\psi \rangle$ of \ref{psidef} are at least delta-function
normalizable. Thus, for almost every $\alpha$ the list of
functions $f_n(\alpha)$ should give the components $(\overline{f}(\alpha))_n$
of an $\alpha$-dependent vector
${\overline f}(\alpha)$
in the Hilbert space $l^2$ of square summable sequences that satisfies
$\int d\alpha |{\overline{f}}(\alpha)|^2= \langle \psi|\psi\rangle$.
On this space,
the coefficients $M_{mn}$ define an antihermitian operator
through $(M {\overline f})_n = \sum_m M_{nm} \overline{f}_m$.  Let
$U(\alpha)$ be the corresponding unitary path ordered exponential ${\cal P}
\exp(-\int_{\alpha_0}^{\alpha}
d \alpha' M(\alpha'))$
satisfying ${{dU^{-1}} \over {d\alpha}} = U^{-1}M$
and let ${\overline f} = U {\overline g}$ for some
${\overline g}$.  Then we have
\begin{equation}
\label{M conn}
({{{\partial} \over {\partial \alpha}}}
f_n + \sum_m M_{nm} f_m) = (U {{{\partial} \over {\partial \alpha}}}
{\overline g})_n
\end{equation}
and we may write \ref{expand} as
\begin{equation}
H | \psi \rangle = \int d \alpha \sum_n   \Bigl( \hbar^2
\bigl( U {{{\partial ^2 } \over {\partial \alpha^2}}}
{\overline g} \bigr)_n + (U {\overline g})_n
E_n(\alpha) \Bigr) |n; \alpha \rangle.
\end{equation}
It follows that $|\psi\rangle$ is an eigenvector of $H$
with eigenvalue $E$ exactly when the corresponding
$\overline{g}$ is a generalized $L^2$ function and satisfies
\begin{equation}
\label{geqn}
\hbar^2{{\partial^2 } \over {\partial \alpha^2}}
\overline{g} + (\tilde{E}(\alpha)\circ \overline{g}) = E \overline{g}.
\end{equation}
Here, $\tilde{E}(\alpha) = U^{-1}(\alpha)\overline{\overline{E}}(\alpha)
U(\alpha)$ and $\overline{\overline{E}}(\alpha)$ is the operator on
$l^2$ such that $(\overline{\overline{E}}(\alpha) \overline f)_n
= E_n(\alpha) \overline{f}_n$.
We will show that for $E \leq 0$, there is one such solution for
every $\overline{g}_0 \in l^2$.

The equation \ref{geqn} to be solved for $\overline{g}(\alpha)$
is essentially a one-dimensional
time independent Schr\"odinger equation for a multiple component
wavefunction with potential $-\tilde{E}(\alpha)$
and energy $-E$.
Thus, existence, normalizability, and multiplicity of the solutions to
$\ref{geqn}$ is related to whether this ``potential"
confines a (fictitious) particle to finite values of $\alpha$ or lets such a
particle escape to the asymptotic regions.  Note that,
as the potential $V^*(\beta^*;\alpha)$
of $H^*(\alpha)$ becomes more and more shallow for large negative
$\alpha$, $E_n(\alpha) \rightarrow 0$ as $\alpha \rightarrow - \infty$
and our fictitious particle is unbound in this direction for $E \leq 0$.
In particular,
for large negative $\alpha$, even the ground state
energy of $H^*(\alpha)$ is positive so that $-E_n(\alpha)$ approaches $0$
from below as $\alpha \rightarrow -\infty$ and there can be no
normalizable eigenvector with $E\leq 0$;
{\em all} solutions to \ref{geqn} with $E \leq 0$ are
delta-function normalizable  in the region $\alpha \leq 0$.

Solutions of \ref{geqn} with $E > 0$ are difficult to analyze in the
$\alpha \rightarrow - \infty$ limit without a uniform bound on the
rate at which $E_n(\alpha)$ vanishes.  Note, however, that the
existence of such a unifofm bound would show that the spectrum of $H$
is discrete for $E > 0$ so that zero is the largest continuum eigenvalue
in dramatic contrast to the separable cases of \cite{QORD}.  A WKB
approximation of the type used below for $\alpha \rightarrow +\infty$
might provide insight but is even more difficult to control than at
the positive $\alpha$ end.  As we only require knowledge of the
spectrum of $H$ on some $[a,b]$ containing zero, we take $a < 0$ and
$b = 0$ and do not concern ourselves with the case $E > 0$.

The remaining task is to study the behavior of $E_n(\alpha)$ for large,
positive $\alpha$.  To do so, we
write $\overline{g}$ in the form
\begin{equation}
\o{g} = {\cal P}\exp[\int_0^\alpha d\alpha' N(\alpha')] \o{g}_0
\end{equation}
where $V = {\cal P}\exp[\int_0^\alpha d\alpha' N(\alpha')]$
is the path ordered exponential that satisfies ${{\partial V}
\over {\partial \alpha}} = NV$.  Our equation \ref{geqn} can now be
written as
\begin{equation}
[N^2 V + \dot{N} V + \tilde{E} V]\o{g}_0 = EV\o{g}_0
\end{equation}
where $\dot{N} = dN/d\alpha$.
In order to show that a solution exists for all $\overline{g}_0$,
we will (approximately) solve the matrix equation
\begin{equation}
\label{mat eq}
N^2 V + \dot{N} V + \tilde{E}V - E V = 0
\end{equation}
As in the finite dimensional WKB method, assume that $N^2 >> \dot{N}$,
so that \ref{mat eq} becomes $N^2 = E\openone -
\tilde{E}$.  Since $\tilde{E}$
is a positive self-adjoint operator with known spectrum,  a number
of square-roots $\sqrt{E \openone - \tilde{E}}$ are easily defined.
For the moment, let $N$ be any one of these.

Now, recall that $\tilde{E}(\alpha) = U^{-1}(\alpha)\overline
{\overline{E}}(\alpha) U(\alpha)$.  It follows that
$N = U^{-1}(\alpha) (\sqrt{E \openone - \o{\o{E}}}) U(\alpha)$ and
we have
\begin{equation}
{{\partial } \over {\partial \alpha }}(UV) -
[{{\partial U} \over {\partial \alpha}}
U^{-1}] UV = U {{\partial V } \over {\partial \alpha}} = \sqrt{E
\openone -\o{\o{E}}} UV
\end{equation}
which has the solution $V = U^{-1}{\cal
P}\exp[\int_0^{\alpha}
(\sqrt{E \openone-\o{\o{E}}}) - M]$.  This form of $V$ will be
convenient for the approximations below.

We now reason as follows.
Consider again $H^*(\alpha)$ and the associated eigenvalue problem
\begin{equation}
\label{evp}
E_n \psi_n = - \hbar^2 \Delta \psi_n + e^{4\alpha}(V-1)\psi_n.
\end{equation}
Here, $\Delta$ is the two-dimensional Laplacian and $\psi_n$ is
the wavefunction $ \psi_n = \langle \beta_+,\beta_- |n\rangle$.
Since \ref{evp} may also be written as
\begin{equation}
{{E_n} \over {\hbar^2}} \psi_n = -
\Delta \psi_n + {{e^{4\alpha}} \over {\hbar^2}}(V-1)\psi_n
\end{equation}
the ratio $E_n/\hbar^2$ must depend only on $\hbar$ and $\alpha$
through $e^{4\alpha}/\hbar^2$.
Similarly, we must have $e^{-4\alpha}E_n = f(e^{4\alpha}/\hbar^2)$
for some function $f$ and it follows that $\alpha \rightarrow
\infty$ is essentially just the model's classical
limit\footnote{Thanks to Charles W. Misner for pointing this out.}.
However, since the potential $(V-1)e^{4\alpha}$ has only a single
critical point (which is, in fact, a global minimum) we expect that
as $\hbar \rightarrow 0$ every energy level falls toward this minimum.
Thus, $e^{-4\alpha}E_n \rightarrow -1$ as $\alpha \rightarrow
\infty$.

This will in turn imply that, for large $\alpha$, the $n$th
state is concentrated near $\beta^{\pm} = 0$.
Since $E_n = \langle n|H^*|n \rangle$, $- \Delta \geq 0$,
and $(V-1) \geq -1$ as $\alpha \rightarrow \infty$,
the behavior of $E_n(\alpha)$ implies that
$\langle n| (V-1)|n  \rangle \rightarrow -1$.   But
$(V-1) = -1$ only at the origin, so
we must have
$\int_{{\bf R}^2 - B_\epsilon} |\psi_n|^2 \rightarrow 0$ for every
$\epsilon$-ball $B_\epsilon$ centered at $\beta^{\pm} = 0$.

For small enough $\epsilon$, the potential $V-1$ within
$B_{\epsilon}$ is essentially that of a (rotationally symmetric)
Harmonic oscillator
($V \approx 8 e^{4\alpha}[\beta^{+2} + \beta^{-2}])$.  Thus, since
$E_n \rightarrow -e^{4\alpha}$ for every $n$, for each $N>0$ there is
some $\alpha$ for which $H^*(\alpha)$ may be replaced by the
rotationally invariant Harmonic
oscillator Hamiltonian $- \Delta + 8 e^{4 \alpha} (\beta_+^2 + \beta_-^2)
-e^{4\alpha}$
when considering states with $n \leq N$ to some accuracy $\epsilon_0$.
At this point, it is convenient to replace the label $n$ with the
two occupation numbers $n_1,n_2$ of the oscillator so that
$E_{n_1,n_2}(\alpha) \approx - e^{4\alpha} + 2 \sqrt{2}e^{2\alpha}\hbar
(n_1 + n_2 +  1)$. Similarly, we find $\langle m_1,m_2| (V-1) |n_1,n_2 \rangle
\approx c_{(m_1,m_2)(n_1,n_2)} e^{-2\alpha}$, for
$c_{(m_1,m_2)(n_1,n_2)}$
independent of $\alpha$ and
vanishing for $|m_1-n_1| > 2$ or $|m_2 - n_2| > 2$, so that
\begin{equation}
M_{(m_1,m_2)(n_1,n_2)} \approx { {e ^{2\alpha} c_{(m_1,m_2)(n_1,n_2)}}
\over {e^{2\alpha}\hbar(m_1+m_2 - n_1 -n_2) }}
\end{equation}
which is also independent of $\alpha$ in this regime.  On the other hand,
$\sqrt{E \openone - \o{\o{E}}} \rightarrow u e^{2\alpha}$ for some
$u^2 = \openone$ so that
we may neglect $M$ in comparison to this term and write
\begin{equation}
\label{sol}
V \approx U^{-1} \exp(u e^{2\alpha}).
\end{equation}
Note in particular
that $N^2 \sim e^{4\alpha}$
while $\dot{N} \sim e^{2\alpha}$ so that for large $\alpha$ we do indeed have
$N^2 >> \dot{N}$.
\subsection{Eigenvectors and the Physical Hilbert Space}
\label{Hphys}

While \ref{sol} (approximately)
solves \ref{mat eq}, the question arises of which
solutions  $V(\alpha)\overline{g}_0$ are (delta-function) normalizable.
Suppose first that $u$
has some eigenvector $\o{g}_0$ with eigenvalue $+1$.  Then, the
corresponding solution $\o{g} = W\o{g}_0$ has diverging $l^2$
norm ($|\o{g}|^2$) as $\alpha \rightarrow \infty$.  However, for any
eigenvector $\o{g}_0$ of $u$ with eigenvalue $-1$, the
corresponding $\o{g} = W \o{g}_0$
is {\it square integrable} on the region $\alpha \geq 0$
($\int_{\alpha \geq0} |\overline{g}|^2 < \infty$) for all square
summable $\overline{g}_0$.  We thus take
$u=-1$ for the general normalizable solution.

For large negative $\alpha$,
we already know that the
behavior of our solutions is that of a free particle.  It
therefore follows that there is one
delta-function normalizable solution to \ref{geqn} for each
$\overline{g}_0 \in l^2$ and
every $E \leq 0$  so that the continuous spectrum of $H$
includes the interval $(-\infty,0]$.
This verifies that $H$ is of the
type described in section \ref{prelim} and provides
one of the labels ($E$) of our useful
basis of states.

In fact, we have shown that all states in $\tilde{\cal H}$ are of the
form
\begin{equation}
|\psi[\overline{g}_0(E)]\rangle = \int dE |\overline{g}_0(E),E\rangle
= \int dE \sum_n (\overline{g}_0(E))_n |\overline{\chi}_n, E\rangle
\end{equation}
where $|\overline{g}_0,E\rangle$ is the solution of $H|\psi\rangle
= E |\psi\rangle$ given by $\overline{g}(\alpha) = V_E(\alpha)
\overline{g}_0$ for $\overline{g}_0 \in l^2$ and $(\overline{\chi}_n)_m
= \delta_{n,m}$.  If, for convenience, we now refer to the states
$|\chi_n,E\rangle$ as $|n,E\rangle$ then we have found a basis for
$\tilde{\cal H}$ such that
$H|n,E\rangle = E|n,E\rangle$ and $\langle n,E|n',E'\rangle
= \delta(E-E') \delta_{n,n'}$.
The physical Hilbert space ${\cal H}_{phys}$ may now
be constructed as in \cite{QORD} as the closure of the span
of $\{|n,0\rangle \}$ with respect to the inner product
$(|n,0\rangle, |m,0\rangle)_{phys} = \delta_{n,m}$.

\subsection{Physical Operators}
\label{ops}

Having constructed ${\cal H}_{phys}$, we now address both
the formation of physical operators on this
space as in \cite{QORD} through integration over $t$ and
the verification of recollapsing behavior.
We first construct a complete set of operators that $i$) has
a straightforward physical interpretation in the classical limit,
$ii$) are defined as bilinear forms on the physical space, and
$iii$) satisfy the recollapse criterion \ref{rec} as bilinear forms.
We will then regularize these operators using a new operator $L$
whose physical interpretation is less than clear.  The regularized
operators, however,
form a complete set of {\it bounded} operators on the physical
Hilbert space which satisfy \ref{rec} in the sense of strong
convergence; that is, convergence on a dense set of states\cite{RS}.

Consider then the
objects $[A]_{\alpha = \tau}$ of \ref{obs}.  From \cite{QORD},
the action of these operators on $|n,0\rangle$ may be written as
\begin{equation}
\label{left}
(|n^*,0\rangle, [A]^L_{\alpha = \tau} |n,0\rangle )_{phys}
= 2 \pi i \langle n^*,0|A [H, \theta(\alpha - \tau)]|n,0\rangle
\end{equation}
and
\begin{equation}
\label{right}
(|n^*,0\rangle, [A]^R_{\alpha = \tau} |n,0\rangle )_{phys}
= 2 \pi i \langle n^*,0|[H, \theta(\alpha - \tau)]A|n,0\rangle
\end{equation}

Now, for any bounded operator $B$ on ${\cal H}$, let
\begin{equation}
B_{reg} = {1 \over {H+i}} \theta(\alpha - \tau') B \theta(\alpha - \tau')
{1 \over {H-i}}.
\end{equation}
This $B_{reg}$ is bounded since $H$ is self-adjoint and is
Hermitian when $B$ is.  As in \cite{QORD}, the operator $[B_{reg}]_{\alpha
= \tau}$
has the same classical limit as $[B]_{\alpha = \tau}$ for $\tau' < \tau$
since, when the constraint is satisfied, the factors of ${1 \over
{h \pm i}}$ combine to have no effect and the extra step functions
$\theta(\alpha - \tau')$ are classically irrelevant for $\tau > \tau'$.

To see that the $[B_{reg}]_{\alpha = \tau}$ give a complete set of
operators, note that $B$ may be chosen to be any bounded function of
$\beta^{\pm}$ or of $P_{\beta^{\pm}}$.  By also inserting into $B$
appropriate functions of $1 \pm {\rm sign} (P_{\alpha})$ as in \cite{QORD},
we may construct objects $[B_{reg}]_{\alpha = \tau}$ whose classical
limits give arbitrary (bounded) functions of $\beta^{\pm}$ or
$P_{\beta^{\pm}}$ evaluated at point where $\alpha = \tau$ and either
$P_{\alpha} > 0$ or $P_{\alpha} < 0$.  However, from the result of
\cite{Wald} each classical solution has $\alpha = \tau$ only
twice (if at all), once with $P_{\alpha} > 0$ and once with
$P_{\alpha} < 0$.  Thus, since we consider the $[B_{reg}]_{\alpha = \tau}$
for {\it all} real $\tau$, this is a complete set of operators for the
Bianchi IX model.

Because all of the
solutions to \ref{geqn} that were used to build ${\cal H}_{phys}$
are square-integrable over the region $\alpha \geq 0$ (see
section \ref{Hphys}), the states
$\theta(\alpha - \tau')|n^*,0\rangle$ and $\theta(\alpha - \tau)|n,0
\rangle$ are normalizable in ${\cal H}$.  It follows
that the physical matrix elements $(|n^*,0\rangle,
[B_{reg}]_{\alpha = \tau} |n,0 \rangle )_{phys}$ are well-defined and finite
since, for example,
\begin{equation}
(|n^*,0\rangle, [B_{reg}]^L_{\alpha = \tau} | n,0\rangle)_{phys}
= 2 \pi \langle n^*,0| \theta(\alpha - \tau') \Big[ B \theta(\alpha - \tau')
{H \over {H-i}} \Big] \theta(\alpha - \tau) |n,0\rangle
\end{equation}
and $B \theta(\alpha - \tau)
{H \over {H-i}}$ is bounded on ${\cal H}$.  This means that
integrals defining $[B_{reg}]_{\alpha = \tau}$ converge {\it as
bilinear forms} on ${\cal H}_{phys}$.  Also, since
$| \theta(\alpha - \tau)|n,0\rangle|^2 \rightarrow 0$ as $\tau \rightarrow
\infty$, these objects converge to
zero as forms and verify the classical recollapsing
behavior.

Now, a modification of \ref{lobs} and \ref{robs} leads to objects that
can be proven to be {\it bounded operators} on
${\cal H}_{phys}$.  To determine if some
$[A]_{\alpha = \tau}$ is a well-defined
operator, it must be checked whether
the norms of the states obtained
by acting with $[A]_{\alpha = \tau}$ on physical
states are finite.
Thus, from \ref{left} and \ref{right},
we are interested in sums of the form
\begin{equation}
\label{IX sum}
\sum_{n^*}  | \langle n^*,0;t'|{\cal O}|n,0;t' \rangle|
\end{equation}
for ${\cal O} = [A,H]\theta(\alpha -\tau)$ and ${\cal O} =
\theta(\alpha - \tau) [H,A]$.
Equivalently, we can replace $[A,H]$ in \ref{IX sum}
with $AH$ or $HA$ as appropriate since $H$
vanishes on the physical subspace.

In order to improve the convergence of \ref{IX sum}, we
consider the new ``level operator" $L = L_0 \oplus \openone^{\perp}$
where the direct sum corresponds to the decomposition ${\cal H}
= \tilde{\cal H}  \oplus \tilde{\cal H}^{\perp}$ of ${\cal H}$
into $\tilde{\cal H}$ and its orthogonal complement.  The operator
$\openone^{\perp}$ is the identity operator on $\tilde{\cal H}^{\perp}$
and $L_0$ is defined by
\begin{equation}
\label{L def}
L_0 |n,E\rangle = n |n,E\rangle
\end{equation}
so that $L$ is self-adjoint.  Now, instead
of \ref{lobs} and \ref{robs}, consider the integrals
\begin{eqnarray}
\label{tildes}
\widetilde{[A]}^L_{\alpha = \tau} &=& \int_{-\infty}^{\infty}
f(L(t)) A(t) {{\partial} \over {\partial t}} \theta(\alpha (t) - \tau)
f(L(t)) \cr
\widetilde{[A]}^R_{\alpha = \tau} &=& \int_{-\infty}^{\infty}
f(L(t)) {{\partial} \over {\partial t}} \theta(\alpha (t) - \tau) A(t)
f(L(t))
\end{eqnarray}
where $f$ is an as yet unspecified function that vanishes for large
$L$.  Here, $L(t) = L$
since $[H,L] = 0$ from \ref{L def}, so that \ref{tildes} amounts
to multiplying expressions \ref{lobs} and \ref{robs} by
$f(L)$ on the left and on the right.  As before,
$\widetilde{[A]}^L_{\alpha = \tau} {}^{\dagger} =
\widetilde{[A]}^R_{\alpha = \tau}$ for self-adjoint $A$,
$\widetilde{[A]}_{\alpha = \tau}  \equiv {1 \over 2}
(\widetilde{[A]}^L_{\alpha = \tau} +
\widetilde{[A]}^R_{\alpha = \tau})$, and
\begin{eqnarray}
\label{IX action}
(|n^*,0\rangle, \widetilde{[A]}^L_{\alpha = \tau}
|n,0\rangle)_{phys} &=& -i 2 \pi
\langle n^*,0| f(L) A H \theta(\alpha - \tau) f(L)
|n,0\rangle \cr
(|n^*,0\rangle, \widetilde{[A]}^R_{\alpha = \tau}
|n,0\rangle)_{phys} &=& i 2 \pi
\langle n^*,0| f(L) \theta(\alpha - \tau) H A
f(L)| n,0\rangle
\end{eqnarray}

Again, let $B_{reg} = {1 \over {H+i}}
\theta(\alpha - \tau') B \theta(\alpha - \tau') {1 \over {H-i}}$.
For this case, the matrix
elements in \ref{IX action} may be written
\begin{equation}
\label{mat elm}
(|n^*,0\rangle, \widetilde{[B_{reg}]}^L_{\alpha  = \tau} |n,0\rangle )_{phys}
= - 2 \pi f(n^*) f(n)
\langle \psi_{n^*}(\tau')| B \theta(\alpha - \tau') {H \over {H-i}}
|\psi_n(\tau)\rangle
\end{equation}
where $|\psi_n(\tau) \rangle = \theta(\alpha(t') - \tau)|n,0
\rangle$ is a state in ${\cal H}$ with finite norm $z_{\tau}(n)$.
Let us choose $f$ such that $|f(n)| \leq { 1 \over {nz_{\tau'}(n) }}$
and $\tau'$ such that $\tau' < \tau$.  Thus, $f$ may depend on
$\tau'$ but not on $\tau$.  Since $z_{\tau}(n)$ is a
nonincreasing function of $\tau$, we have
$z_{\tau}(n) \leq z_{\tau'}(n)$, and
the matrix elements \ref{mat elm} are bounded in
absolute value by
\begin{equation}
2\pi \ |f(n) f(n^*)| \
z_{\tau}(n)z_{\tau'}(n^*)||B|| \leq {{||B||} \over {nn^*}}
{{z_{\tau}(n)} \over {z_{\tau'}(n)}} 2\pi
\end{equation}
where $||B||$ is the operator norm of $B$ on ${\cal H}$.

The physical norms of $[A]^{L,R}_{\alpha = \tau} |n,0\rangle$
are then bounded by
\begin{eqnarray}
\label{n bounds}
||\widetilde{[B_{reg}]}^L_{\alpha = \tau} |n,0\rangle ||^2_{phys}
&\leq& {{z_{\tau}(n)} \over {z_{\tau'}(n)}} \sum_{n^*} {1 \over
{(n^*)^2}} (2\pi)^2\cr
||\widetilde{[B_{reg}]}^R_{\alpha = \tau} |n,0\rangle ||^2_{phys}
&\leq& \sum_{n^*} {1 \over {(n^*)^2}} {{z_{\tau}(n^*)} \over
{z_{\tau'}(n^*)}} (2\pi)^2
\end{eqnarray}
and $\widetilde{[B_{reg}]}_{\alpha = \tau}$ is a {\it bounded}
operator on ${\cal H}_{phys}$.  Furthermore, the action of
$\widetilde{[B_{reg}]}_{\alpha = \tau}$ on any of the states $|n,0\rangle$
vanishes as $\tau \rightarrow \infty$ since both bounds in
\ref{n bounds} become zero in this limit.  This is clear for the first
bound and follows for the second from the facts that the sum over $n^*$
converges and that $z_{\tau}(n^*)$ is a positive nonincreasing function of
$\tau$ that vanishes as $\tau \rightarrow \infty$.  It follows
that $\widetilde{[B_{reg}]}_{\alpha = \tau}$ converges strongly
to zero and, in fact, $\widetilde{[B_{reg}]}^R_{\alpha = \tau}$ converges
uniformly to zero.

Again, in the classical limit and on the constraint surface,
the factors of ${1 \over {h \pm i}}$ cancel
and the factors of
$\theta(\alpha - \tau')$ have no effect for $\tau' \leq \tau$.
In addition, since $[L,H] = 0$, $L$ is conserved
along a classical solution and,
for a given solution $s$, $[f(l)
{1 \over {h+i}} \theta(\alpha - \tau')
b \theta(\alpha - \tau') {1 \over {h-i}} f(l)]_{\alpha = \tau}(s)$
tends to zero for large $\tau$ if and only if
$[b]_{\alpha = \tau} \rightarrow
0$ as $\tau \rightarrow \infty$.  We
thus conclude that our model describes a quantum theory of
recollapsing cosmologies.

Since $l$ is some function on the phase space, it can be
expressed in terms of $\beta^{\pm}$ and $p_{\beta^{\pm}}$.  But $b$ may
be an arbitrary bounded function of $\beta^{\pm}$ and $p_{\beta^{\pm}}$
so that $\{[f(l)]^2b\}$ is still sufficient to separate points on the phase
space.  In this way, we have constructed a complete set of quantum
operators for the Bianchi IX minisuperspace.

\section{Discussion}
\label{disc}

As claimed in the introduction, we have
applied the techniques of \cite{QORD} to the Bianchi IX
cosmological model and thereby construct a quantum theory.
Specifically, a Hilbert space structure on the generalized
eigenvalues of $H$ was induced by the
auxiliary space ${\cal H}_{aux}$,
a complete set of observables was constructed, and
recollapsing behavior was derived, all despite the complicated and
perhaps chaotic nature of the model.

Finally, since the literature on the mixmaster cosmology is extensive, a short
comparison of section \ref{IX} with previous work is presented below.
Note first that the observables studied here are, in spirit, much like those
of \cite{SL} (and suggested by Moncrief), which describe anisotropies
and momenta at the maximal value of $\alpha$.  Such objects are easily
written down in the formalism used here by choosing $Z$ in
\ref{obs} and \ref{robs} to be $P_{\alpha}$ and $\tau$ to be zero, since
$P_{\alpha}$ vanishes classically on the maximal volume surface.
Note that \cite{SL} also defines observables away from
maximal volume which, for volumes suitably close to maximal are (exactly)
unitarily related.  This ``unitary evolution" then breaks down at a finite
separation from maximal volume. In our case, because the physical
operators $[A]_{\alpha = \tau}$ vanish for large $\tau$, we also find that
the $[A]_{\alpha = \tau}$ and $[A]_{\alpha = \tau'}$ are not unitarily
related when the separation between $\tau$ and $\tau'$ is large.
On the other hand, there is no
reason to think that {\it any} of the $[A]_{\alpha = \tau}$ of
\ref{IX} are unitarily related or that there should be a preferred
finite separation at which such a unitary
evolution breaks down.  Thus, we expect these quantizations
to be inequivalent.

We now single out the recent construction of exact solutions to the
quantum constraint by Moncrief and Ryan \cite{Mon} for more detailed
discussion.  Such solutions
were found earlier by Kodama \cite{Kodama}
for the constraint written in terms
of Ashtekar variables, but the work of \cite{Mon} uses the variables
of section \ref{IX} so that the comparison with our approach is
direct.  A similar comment applies to the super-symmetric solutions of
Graham \cite{graham}.

However, the quantum states of \cite{Mon} do not solve the
constraint \ref{IX constr} as they are related to a different
factor ordering of the classical expression.  Instead, they solve
\begin{equation}
0 = H'\psi = {{\partial^2} \over {\partial \alpha^2}} \psi - B
{{\partial} \over {\partial \alpha}} \psi - {{\partial}^2 \over
{(\partial \beta^+)^2}}\psi - {{\partial}^2 \over {(\partial \beta^-)^2}}
\psi + e^{4 \alpha}(V-1)\psi
\end{equation}
for some real $B$ where, following \cite{Mon}, we have dropped
the $\hbar$'s.

Thus, our setting does not coincide with that of \cite{Mon} and
some effort will be needed to reconcile them.  In particular,
\cite{Mon} does not consider an auxiliary  Hilbert space and we
must introduce one here for comparison with section \ref{IX}.  Note that
\cite{Mon} provides explicit solutions only for $B = -6 \neq 0$
and that for $B \in {\bf R}$, $B \neq 0$, the differential
operator above is not Hermitian in the Hilbert space
$L^2({\bf R}^3,d\alpha d\beta^+d\beta^-)$.
This operator is, however, self-adjoint in the Hilbert space
$L^2({\bf R}^3, e^{-B\alpha} d\alpha d\beta^+ d\beta^-)$, which
we will therefore take to be our auxiliary space.

A rescaling of the wavefunction $\psi = e^{B\alpha/2}\phi$
leads to the equivalent constraint
\begin{equation}
\label{BC}
0 = H'' \phi =  {{\partial^2} \over {\partial \alpha^2}} \phi - {{B^2}
\over 4} \phi
- {{\partial}^2 \over
{(\partial \beta^+)^2}} \phi - {{\partial}^2 \over {(\partial \beta^-)^2}}
\phi + e^{4 \alpha}(V-1)\phi
\end{equation}
on $\phi \in L^2({\bf R}^3, d^3x)$.
Note that the arguments of \ref{IX} apply as well to \ref{BC} as
to \ref{IX constr}, the only difference being a relative shift of
$E_n(\alpha)$ by ${{B^2} \over 4}$.  Thus, the behavior at large
$\alpha$ is unchanged and all energies are ``bound" at this end.
The difference appears at large negative $\alpha$, where the
behavior of our wavefunction reduces to that of
a free particle.  If $H''\phi = E\phi$, then this
free particle has energy $-(E+ {{B^2} \over 4})$.
Thus, the (probable) upper bound on the continuous spectrum moves from
zero to $- {{B^2} \over 4}$.
In particular, the solutions described by \cite{Mon}
vanish rapidly as $\alpha \rightarrow + \infty$ but become
asymptotically constant as $\alpha \rightarrow - \infty$.
The rescaled wavefunction $\phi$ then behaves as $e^{-B\alpha/2}$
for large negative $\alpha$ so that such solutions are normalizable
whenever $B < 0$ such as when $B= -6$.

Thus, zero is now in the
{\it discrete} spectrum of the constraint and the construction of
\ref{Hphys} is unnecessary since the relevant eigenvectors
form a subspace of the auxiliary space and so inherit the inner
product directly.  Also, because the spectrum is discrete,
from \cite{QORD} we expect that the
integrals \ref{lobs} and \ref{robs} which attempt to define our
gauge-invariant operators will now fail to converge.

Finally, if we allow $B$ to be imaginary then the original
operator $H'$ {\it is} self-adjoint on $L^2({\bf R}^2, d^3x)$.
The same analysis follows as for $B$ real, but now the boundary
between the discrete and continuous spectrum moves to $E = |B|^2/4 > 0$
so that zero again lies in the continuous spectrum.  Solutions of
the form described by \cite{Mon} would then correspond to elements
of ${\cal H}_{phys}$ of \ref{IX} and the construction of gauge-invariant
operators proceeds as in \ref{IX}.

\acknowledgements

The author would
also like to express his thanks to Abhay Ashtekar,
Jos\'e Mour\~ao,
Charles W. Misner, Seth Major, Alan Rendall,
and Lee Smolin for helpful discussions.
Thanks also go
to Grigori Rozenblioum and sci.math.research for help in locating
reference \cite{QM} and to
Jos\'e Mour\~ao for help with its translation.
This work was supported by NSF grant PHY 93-96246 and
by funds provided by The Pennsylvania State University.

\appendix
\section{Finiteness of the region $V < E$}
\label{finite}

We now quickly derive the fact that the area $A(E)$
of the region of the $\beta^+,\beta^-$ plane on which $V$ of
\ref{V} is less than $E$ is finite for any $E$ in ${\bf R}$.
To do so, we use the $2\pi/3$ discrete rotational symmetry of $V$ and its
well-known ``triangular" shape \cite{MTW}.  Recall that this
``triangle" points along the $\beta^+$ axis.

For large positive $\beta^+$, let $h^E(\beta_+)$ be the value of
$\beta_-$ for which $V(\beta_+,\beta_-) = E$.  In this part of the
plane, there are
always two solutions $\pm h^E(\beta_+)$ and we
take $h^E(\beta_+)$ to be the positive one.  $A(E)$ is then finite
if we have
\begin{equation}
\label{hint}
\int_{\lambda}^{\infty} h^E(\beta_+) d\beta_+ < \infty
\end{equation}
for any finite $\lambda$.

{}From \ref{V}, it follows that
\begin{equation}
V(\beta_+,\beta_-) < - {4 \over 3} e^{-2 \beta_+}
+ {2 \over 3} e^{4 \beta_+} (\cosh 4 \sqrt{3} \beta_- - 1)
\end{equation}
so that $h^E(\beta_+)$ is less than the $h_0^E(\beta_+)$ defined by
\begin{equation}
0 = -2 e^{-2\beta_+} + e^{4 \beta_+} (\cosh 4 \sqrt{3} h_0^E(\beta_+) -1)
\end{equation}
Clearly, $h_0^E \rightarrow 0$ for large $\beta_+$ so that, if
$\beta_+$ is large enough, $\cosh 4\sqrt{3} h_0^E -1 \geq {1 \over 2}
[ 3(2^3)(h_0^E)^2 ]$.  It follows that
$h^E \leq h_0^E \leq {1 \over 6} e^{-6\beta_+}$
and that \ref{hint} holds.

\end{document}